\pdfoutput=1
\documentclass[aps,rmp,twocolumn,groupedaddress,amsmath,floatfix]{revtex4-2}
\usepackage{graphicx}

\usepackage[english]{babel}
\usepackage[LGR,T1]{fontenc}
\usepackage[utf8]{inputenc}
\usepackage[scaled=.95,helvratio=.96]{newtxtext}
\usepackage[varqu,varl]{zi4}
\usepackage{cabin}
\usepackage[vvarbb]{newtxmath}
\DeclareMathAlphabet{\mathpzc}{OT1}{pzc}{m}{it}
\newcommand{\layerE}{\alpha} 
\newcommand{\layerF}{\beta} 

\usepackage{subcaption}
\usepackage{ragged2e}
\DeclareCaptionJustification{justified}{\justifying}
\addto\captionsenglish{}

\usepackage[sf,bf,small,noindentafter]{titlesec}
\titleformat*{\section}{\normalsize\bfseries\sffamily}
\titlespacing*{\section}{0pt}{1em}{0em}
\titleformat{\subsection}[runin]{\bfseries}{}{}{}
\titlespacing{\subsection}{0pt}{1em}{0pt}

\usepackage[activate={true,nocompatibility},final,tracking=true,kerning=true,spacing=false,factor=1100,stretch=10,shrink=10]{microtype}
\microtypecontext{spacing=nonfrench}
\bibpunct{}{}{,}{s}{,}{,}

\usepackage{booktabs}

\newcommand{\mytoprule}{\specialrule{0.1em}{0em}{0em}}
\newcommand{\mybottomrule}{\specialrule{0.1em}{0em}{0em}} 
\newcommand{\mymidrule}{\specialrule{0.05em}{0em}{0em}}
\usepackage{float}
\usepackage{threeparttable}

\usepackage[hidelinks]{hyperref}

\usepackage{siunitx}

\begin{document}
\makeatletter
\renewcommand\@biblabel[1]{#1.}
\makeatother

\newcommand*{\citen}[1]{%
  \begingroup
    \romannumeral-`\x 
    \setcitestyle{numbers}%
    \cite{#1}%
  \endgroup   
}


\title{Mapping flows on hypergraphs} 


\author{Anton Eriksson}
\email[]{anton.eriksson@umu.se}
\affiliation{Integrated Science Lab, Department of Physics, Ume{\aa} University, SE-901 87 Ume{\aa}, Sweden}

\author{Daniel Edler}
\affiliation{Integrated Science Lab, Department of Physics, Ume{\aa} University, SE-901 87 Ume{\aa}, Sweden}

\author{Alexis Rojas}
\affiliation{Integrated Science Lab, Department of Physics, Ume{\aa} University, SE-901 87 Ume{\aa}, Sweden}

\author{Martin Rosvall}
\affiliation{Integrated Science Lab, Department of Physics, Ume{\aa} University, SE-901 87 Ume{\aa}, Sweden}


\date{December 23, 2020} 

\begin{abstract}
Hypergraphs offer an explicit formalism to describe multibody interactions in complex systems.
To connect dynamics and function in systems with these higher-order interactions, network scientists have generalised random-walk models to hypergraphs and studied the multibody effects on flow-based centrality measures.
But mapping the large-scale structure of those flows requires effective community detection methods.
We derive unipartite, bipartite, and multilayer network representations of hypergraph flows and explore how they and the underlying random-walk model change the number, size, depth, and overlap of identified multilevel communities.
These results help researchers choose the appropriate modelling approach when mapping flows on hypergraphs. 
\end{abstract}

\maketitle

Researchers model and map flows on networks to identify important nodes and detect significant communities\cite{brin1998anatomy,simonsen2004diffusion,rosvall2008maps,delvenne2010stability}. From small to large system scales, random walk-based methods help to uncover the inner workings of the systems the networks represent~\cite{boccaletti2006complex,fortunato2010community}. When standard network models fail to adequately represent a system's interactions, researchers turn to higher-order models of complex systems~\cite{lambiotte2019networks,battiston2020networks}, including multilayer networks~\cite{mucha2010community,kivela2014multilayer,de2016physics} for multitype interactions, non-Markovian networks~\cite{rosvall2014memory,scholtes2014causality,xu2016representing} for multistep interactions, and combinatorial models such as simplicial complexes~\cite{parzanchevski2017simplicial,salnikov2018simplicial,iacopini2019simplicial,schaub2020random} and hypergraphs~\cite{zhou2007learning,chitra2019random,carletti2020random,carletti2020randomwalks} with nodes in hyperedges for multibody interactions.

While several methods can identify flow-based communities in multilayer~\cite{mucha2010community,de2015identifying,jeub2017local} and memory~\cite{rosvall2014memory,scholtes2014causality,xu2016representing} networks with non-Markovian dynamics,
researchers have just begun to unravel the large-scale systemic effects of multibody interactions captured by hypergraphs~\cite{carletti2020randomwalks}.
However, different systems and research questions call for different random walk and hypergraph models: Random walks can be lazy, able to visit the same node multiple times in a row, or non-lazy and forced to move on.
Hyperedges can have arbitrary weights, and nodes can have hyperedge-dependent weights.
Because these and other models can be represented with different network types -- bipartite, unipartite, and multilayer -- the questions multiply:
How do different hypergraph random-walk models combined with different network representations change the flow dynamics at scales captured by communities?

For example, random walks on hypergraphs can model the flow of ideas in co-authorship networks.
A node represents an author, and a hyperedge connects all authors of a paper. In the simplest dynamics, a random walker on a node picks a random hyperedge among those that contain the node and steps to a random node of the picked hyperedge.
Then repeats.
Excluding author self-links for non-lazy walks or including hyperedge weights from paper citations or using hyperedge-dependent node weights for varying author contributions are natural model variations that generate different dynamics~\cite{carletti2020random,chitra2019random}.
How does the organisation of authors in nested communities from research groups to research areas change with random-walk model and representation?

For lazy random walks on hypergraphs with self-links and hyperedge-independent node weights, random walks on weighted, undirected networks generate equivalent dynamics~\cite{chitra2019random}.
Each hyperedge becomes a clique with properly adjusted link weights.
This projection enables standard flow-based methods developed for weighted networks to identify communities where random walks stay for a long time.
Non-lazy walks or walks with hyperedge-dependent node weights require directed networks~\cite{chitra2019random}.
A bipartite representation provides hyperedge assignments, and a multilayer representation enables overlapping communities.

Representing hypergraphs with bipartite networks requires weighted, directed links between two sets of nodes: one for the nodes and one for the hyperedges.
Picking a random hyperedge becomes an explicit step to a hyperedge node.
Non-lazy walks on the hypergraph require non-backtracking walks on the bipartite network~\cite{alon2007non}. 
With proper normalisation, the node-visit rates stay the same.
Though unipartite and bipartite representations give identical node flows, the bipartite representation's link flows from nodes to hyperedge nodes and back to nodes can induce more flows between communities and alter the optimal community composition. The community-detection algorithm must also assign more nodes, which implies more degrees of freedom and a larger search space.

\begin{figure*}[!thb]
 \centering
 \includegraphics[width=\textwidth]{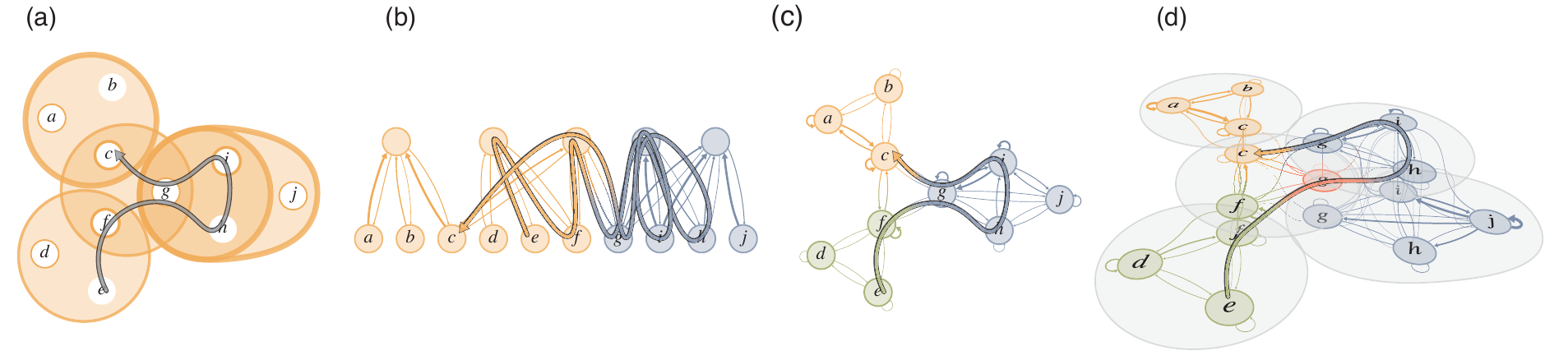}
 \caption{A schematic hypergraph represented with three types of networks. (a) The schematic hypergraph with weighted hyperedges and hyperedge-dependent node weights. Thin borders for weight 1 and thick borders for weight 3. A lazy random walk on the schematic hypergraph represented on: (b) a bipartite network, (c) a unipartite network, and (d) a multilevel network. The colours indicate optimised module assignments, in (d) for hyperedge-similarity walks.
 \label{fig:hypergraphRepresentations}}
\end{figure*}

Multilayer networks represent the hyperedges as layers with fully connected groups of nodes.
Each node is present in each of its hyperedge layers.
Hyperedge weights become layer weights, and hyperedge-dependent node weights become layer-dependent node weights.
Though the node visit rates aggregated over layers remain the same, multilayer networks multiply the degrees of freedom and enable new models.
Reducing the inter-layer link weights increases the time a random walker spends within a hyperedge before moving to another.
Reducing the inter-layer link weights only between dissimilar layers reinforces flows within similar layers.
The search space expands when nodes can belong to multiple overlapping communities.
 
The many combinations of random-walk models and representations available to address specific research problems require us to ask, for different data and different questions, which model and representation is best?

To address which combination of model and representation is best for answering different questions about various hypergraph data, we derive unipartite, bipartite, and multilayer network representations of hypergraph flows with identical node-visit rates for the same random-walk model.
For unique node-visit rates when a representation requires directed links, we apply an unrecorded teleportation scheme robust to changes in the teleportation rate and that preserves the node-visit rates when teleportation is superfluous in undirected networks~\cite{lambiotte2012ranking}.
The information-theoretic and flow-based community detection method Infomap~\cite{edler2017mapping}
allows us to explore how different hypergraph random-walk models and network representation change the number, size, depth, and overlap of identified multilevel communities.

By analysing schematic and real hypergraphs, we find that the bipartite network representation requires the fewest links and enables the fastest community detection. A multilayer network representation that reinforces flows within similar layers gives the deepest modular structures with the most overlapping communities but at a high computational cost. The unipartite network representation provides a trade-off between the two, with intermediate compactness, speed, and detectable modular regularities.

\section*{Results and Discussion}
\subsection*{Modelling flows on hypergraphs}.
We model flows on hypergraphs with random walks, using
hypergraphs with nodes $V$, hyperedges $E$ with weights $\omega$, and hyperedge-dependent node weights $\gamma$.
Each hyperedge $e$ has a weight $\omega(e)$.
Each node $u$ with incident hyperedges $E(u) = \{ e \in E : u \in e \}$ has a weight $\gamma_e(u)$ for each incident hyperedge $e$.
To simplify the notation when normalising weights into probabilities, we denote node $u$'s total incident hyperedge weight $d(u) = \sum_{e \in E(u)}\omega(e)$ and hyperedge $e$' total node weight $\delta(e) = \sum_{u \in e}\gamma_e(u)$~\cite{chitra2019random}.
With these weights, a lazy random walker
moves from node $u$ at time $t$ to node $v$ at time $t+1$ in three steps by~\cite{chitra2019random}:
\begin{enumerate}
    \item Picking hyperedge $e$ among node $u$'s hyperedges $E(u)$ with probability $\frac{\omega(e)}{d(u)}$.
    \item Picking one of the hyperedge $e$'s nodes $v$ with probability  $\frac{\gamma_e(v)}{\delta(e)}$.
    \item Moving to node $v$.
\end{enumerate} 
Variations include non-lazy walks, which never visit the same node twice in a row with a modified second step
\begin{enumerate}\setcounter{enumi}{1}
\item[2b.] Picking one of the hyperedge $e$'s nodes $v \ne u$ with probability  $\frac{\gamma_e(v)}{\delta(e)-\gamma_e(u)}$,
\end{enumerate} 
and teleporting walks, which jump to a random node at some rate to ensure that all nodes can be reached from any node in a finite number of moves, so-called ergodic walks.
We pick the next hyperedge based on its similarity to the previously picked hyperedge in hyperedge-similarity walks, which are useful for modelling flows that tend to stay among similar hyperedges such as among research papers with similar author lists and likely similar topics.
These walks require memory and correspond to a higher-order Markov chain model because they depend on the previously picked hyperedge.

The bipartite, unipartite, and multilayer network representations have different advantages and limitations (Fig.~\ref{fig:hypergraphRepresentations}).
A weighted, undirected network suffices for memoryless lazy random walks without hyperedge-dependent node weights, hyperedge-dependent node weights require directed networks, and hyperedge-similarity walks require multilayer networks. 

\begin{figure}[htp]
    \centering
    \includegraphics[width=\columnwidth]{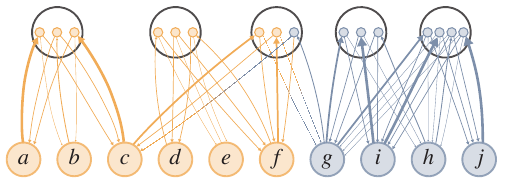}
    \caption{Bipartite network with state nodes for non-lazy random walks. To prevent random walks on bipartite networks from visiting the same node at the bottom twice in a row by backtracking from the hyperedge node at the top, we use state nodes in the hyperedge nodes. Each hyperedge node requires one state node for each node in the hyperedge. The state nodes have one incoming link from its source node and outgoing links to all other nodes in the hyperedge. Colours indicate the optimised partition in Fig.~\ref{fig:schematicalluvial}(b).}
    \label{fig:bipartiteNonBacktracking}
\end{figure}

Bipartite networks offer the most direct representation of the three-step random-walk process above.
We represent the hyperedges with hyperedge nodes, and the three steps become a two-step walk between the nodes at the bottom and the hyperedge nodes at the top in Fig.~\ref{fig:hypergraphRepresentations}b.
For simplicity, we refer to them as nodes and hyperedge nodes.
First a step from a node~$u$ to a hyperedge~node~$e$,
\begin{align}
    P_{ue} = \frac{\omega(e)}{d(u)},
\end{align}
and then a step from the hyperedge node to a node~$v$,
\begin{align}
    P_{ev} = \frac{\gamma_e(v)}{\delta(e)}.
\end{align}
By starting the random walk on the nodes and taking two steps at a time, corresponding to a two-step Markov process~\cite{kheirkhahzadeh2016efficient}, hyperedge nodes are only intermediate stops with zero flow when the random walk is back on the nodes after two steps.
The stationary distribution of the random walk is concentrated to the nodes. 
For non-lazy walks represented with bipartite networks, we use so-called state nodes\cite{edler2017mapping} in the hyperedge nodes.
One state node for each incoming link has out-links to all nodes in the hyperedge, except the incoming link's source ensures that the walks are not backtracking (Fig.~\ref{fig:bipartiteNonBacktracking}). 

To represent the random walk on a unipartite network, we project the three-step random-walk process down to a one-step process between the nodes and describe it with the transition rate matrix
\begin{align}
    P_{uv} = \hspace{-1em}\sum_{e \in E(u,v)}\hspace{-1em} P_{ue}P_{ev} = \hspace{-1em}\sum_{e \in E(u,v)} \frac{\omega(e)}{d(u)}\frac{\gamma_e(v)}{\delta(e)},
\end{align}
where $E(u,v) = \{ e \in E : u \in e, v \in e \}$ is the set of hyperedges incident to both nodes $u$ and $v$.
Each hyperedge forms a fully connected group of nodes (Fig.~\ref{fig:hypergraphRepresentations}c).
Unipartite networks for non-lazy walks have no self-links.
Compared with the bipartite representation, the unipartite representation with fully connected groups of nodes requires more links.

To represent the random walk on a multilayer network, we project the three-step random-walk process down to a one-step process on state nodes in separate layers $\layerE$ for each hyperedge $e$.
A state node $u^{\layerE}$ represents $u$ in each layer $\layerE \in {E}(u)$ that contains the node.
All state nodes in the same layer form a fully connected set (Fig.~\ref{fig:hypergraphRepresentations}d).
The transition rate between state node $u^{\layerE}$ in layer $\layerE$ and state node $v^{\layerF}$ in layer $\layerF$ is
\begin{align}
    P_{uv}^{\layerE\layerF} = \frac{\omega(\layerF)}{d(u)}\frac{\gamma_{\layerF}(v)}{\delta(\layerF)}\text{ for } \layerF \in E(u,v).
\end{align}
Node $u$'s state node visit rates in different layers sum to $u$'s visit rate in the unipartite and bipartite representations. 
With one state node per hyperedge layer that contains the node, the multilayer representation requires the most nodes and links to describe the walk.
But this cost comes with benefits: 
the multilayer representation can describe higher-order Markov chains, which can capture more regularities in the data.

For example, a useful variant of the basic hypergraph random walk is to pick a hyperedge not only proportional to its weight but also proportional to its similarity to the hyperedge picked in the previous step.
To include hyperedge-dependent node weight information in the similarity measure, we use one minus the Jensen-Shannon divergence (JSD) between the transition rate vectors $\mathbf{P}_{\layerE v}$ and $\mathbf{P}_{\layerF v}$ to nodes at layers $\layerE$ and $\layerF$ 
as the hyperedge coupling strength,
\begin{align}
    D_{u}^{\layerE\layerF} &= \omega(\layerF)\left[1 - JSD(\layerE,\layerF)\right] \nonumber\\
    &=\omega(\layerF)\biggl[1 - H\left(\frac{1}{2}\mathbf{P}_{\layerE v}+\frac{1}{2}\mathbf{P}_{\layerF v}\right)\nonumber \\
    &\phantom{=\omega(\layerF)\biggl[}+\frac{1}{2}H\left(\mathbf{P}_{\layerE v}\right) +\frac{1}{2}H\left(\mathbf{P}_{\layerF v}\right)\biggr]
\end{align}
for $\layerF \in E(u,v)$.
With node~$u$'s total incident hyperedge weight in layer~$\layerE$
\begin{align}
    S_{u}^{\layerE} = \sum_{\layerF \in E(u)} D_{u}^{\layerE\layerF},
\end{align}
the hyperedge-similarity walk has the transition rates
\begin{align}
    P_{uv}^{\layerE\layerF} = \frac{D_{u}^{\layerE\layerF}}{S_{u}^{\layerE}}\frac{\gamma_{\layerF}(v)}{\delta(\layerF)}\text{ for } \layerF \in E(u,v).
\end{align}
Because the transition rates at a node depend on the current layer, the random walks generate non-Markovian dynamics that a unipartite or bipartite network representation cannot capture.


To ensure ergodic node-visit rates, we derived an unrecorded teleportation scheme that leaves the node-visit rates unchanged when teleportation is superfluous for hypergraphs with hyperedge-independent node weights, robust to changes in the teleportation rate when teleportation is needed\cite{lambiotte2012ranking}, and independent of the representation (see Methods).

\bigskip\subsection*{Mapping flows on hypergraphs}.
To identify flow-based communities or modules in hypergraphs,
we seek to compress a modular description of random walks on the network representations guided by their links.
We cast the problem of finding flow-based communities in hypergraphs as a minimum-description-length problem with the map equation framework~\cite{rosvall2008maps}. With this compression-based framework, we can compare how much the different representations compress modular flows.

When used to detect communities, the representation matters because bipartite, unipartite, and multilayer networks provide the community-detection algorithm Infomap with different degrees of freedom~\cite{edler2017mapping}.
Infomap assigns only nodes to communities in a unipartite network, but assigns also hyperedge nodes in a bipartite network.
The multilayer network, with a state node for each hyperedge a node belongs to, implies even more node assignments and possibly overlapping communities.

\begin{figure}[htb]
 \centering
    \includegraphics[width=0.95\columnwidth]{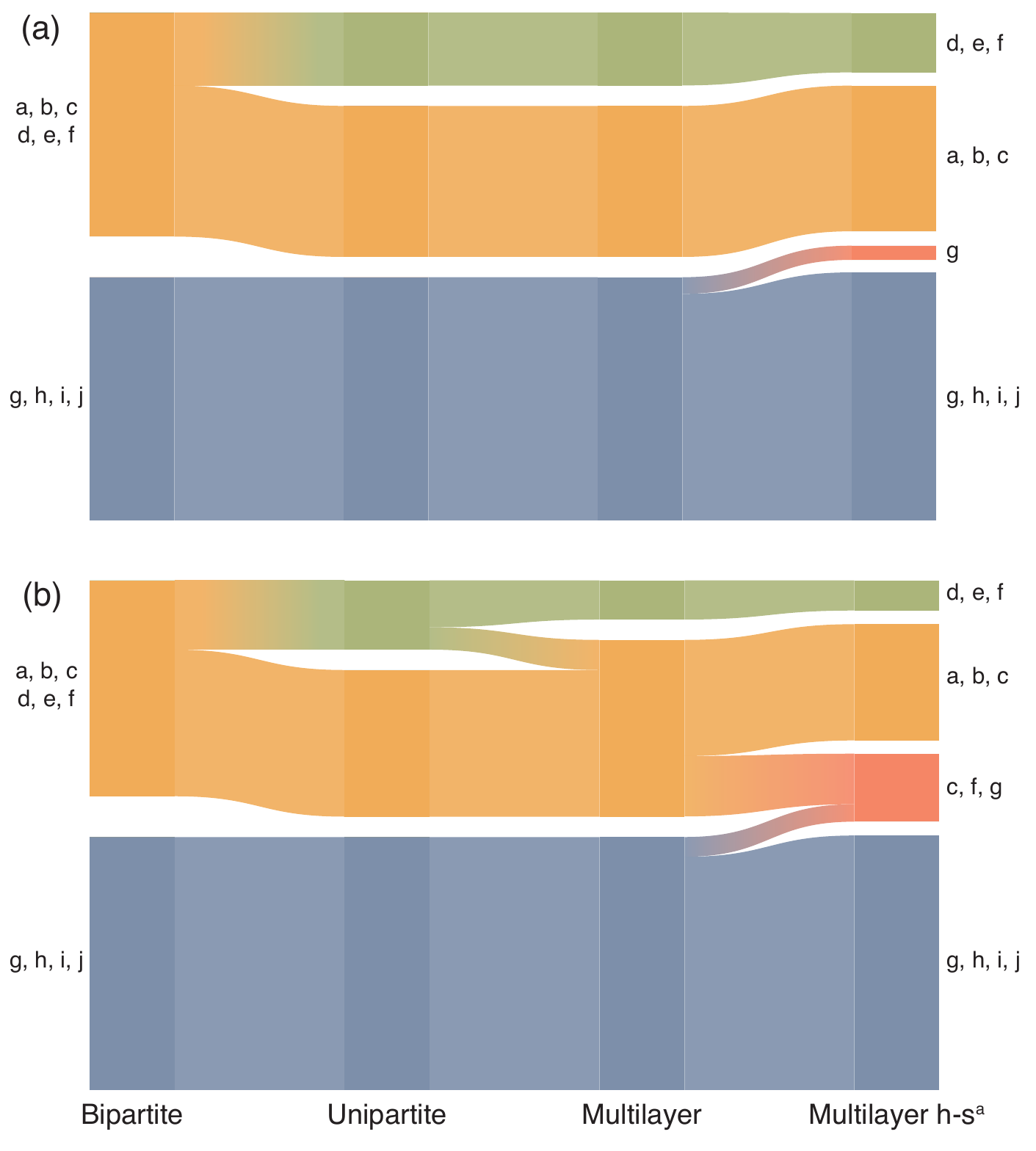}
    \caption{Alluvial diagrams of optimal partitions for the schematic hypergraph in Fig.~\ref{fig:hypergraphRepresentations}. (a) Optimal partitions for lazy walks represented with the networks in Fig.~\ref{fig:hypergraphRepresentations}(b-d). (b) Optimal partitions for non-lazy walks.
    \label{fig:schematicalluvial}}
\end{figure}

When mapping flows modelled by lazy and non-lazy random walks on the schematic network in Fig.~\ref{fig:hypergraphRepresentations}, the optimal partitions of the bipartite networks have two communities, whereas the unipartite and multilayer networks have three communities (Table~\ref{table:schematic} and Fig.~\ref{fig:schematicalluvial}). The bipartite network favours fewer modules -- using the optimal three-module partition of the unipartite network on the bipartite network gives code length 3.29 bits instead of 2.90 bits for two modules –– because the random walker transitions more frequently between modules when they include hyperedges:
Even if a hyperedge node contains no flows at the end of each two-step walk from node through hyperedge node to node, assigning it to a module costs extra bits when it has nodes in multiple modules.
For example, if nodes $a$, $b$, and $c$ in the bipartite network in Fig.~\ref{fig:hypergraphRepresentations}(b) would belong to a third green module as in the optimal unipartite solution, and the random walker at node $c$ would return to the hyperedge it comes from before revisiting node $c$, it would first need to exit the green module and enter the orange module, then exit the orange module and re-enter the green module.
The corresponding walk on the unipartite network stays within the green module.
As a result, the unipartite network representation favours more, smaller modules than the bipartite network representation for lazy and non-lazy walks (Table~\ref{table:schematic}). 

\begin{table}[tbp]
\caption{Optimal flow-based communities of the schematic hypergraph in Fig.~\ref{fig:hypergraphRepresentations} represented with different networks. The number of nodes includes state nodes for the multilevel representations and the bipartite non-lazy representation.
We measure the overlap as the perplexity of the optimal solutions (see Methods).
\label{table:schematic}}
\centering
\setlength{\tabcolsep}{4pt}
\begin{small}
\begin{tabular}{@{}lccccc@{}}
\mytoprule\noalign{\smallskip}
Representation   & Nodes & Links & Modules & Codelength & Overlap \\
{ }              & &  &  & (bits) &\\
\noalign{\smallskip}
\mymidrule\noalign{\smallskip}
\footnotesize{\emph{Lazy}} &&&&& \\
\quad Bipartite             & 15 & 32 & 2 & 2.90 & -- \\
\quad Unipartite            & 10 & 40 & 3 & 2.35 & -- \\
\quad Multilayer            & 16 & 98 & 3 & 2.35 & 1.00 \\
\quad Multilayer h-s\footnotemark[1]        & 16 & 98 & 4 & 2.28 & 1.09\\
\noalign{\smallskip} \footnotesize{\emph{Non-lazy}} &&&&&\\ 
\quad Bipartite             & 26 & 52 & 2 & 3.00 & -- \\
\quad Unipartite            & 10 & 30 & 3 & 2.63 & -- \\
\quad Multilayer            & 16 & 68 & 3 & 2.62 & 1.10 \\
\quad Multilayer h-s\footnotemark[1]        & 16 & 68 & 4 & 2.32 & 1.29 \\
\mybottomrule
\end{tabular}
\footnotetext[1]{hyperedge-similarity}
\end{small}
\end{table}

Multilayer networks enable further compression with overlapping modules. But for this small network, only non-lazy walks give overlapping modules with 0.01 bits compression gain (Table~\ref{table:schematic}). With walks that preferentially move to similar hyperedges, the optimal partitions of the multilayer hyperedge-similarity network representations for lazy and non-lazy random walks both have more overlap in four modules  (Table~\ref{table:schematic} and Fig.~\ref{fig:schematicalluvial}). The hyperedge-similarity walks favour these overlapping modules because they stay longer within them than the regular walks.

For a given random-walk model, the representations give equivalent node-visit rates but alter the link flows, and with different link flows, the optimal partition can change. The bipartite network representation favours partitions with fewer modules than the unipartite network representation because assigning hyperedge nodes to modules implies encoding more transitions between modules.
Multilayer representations, especially with walks that spend longer time among similar hyperedges, favour more overlapping modules. The random-walk model determines how much the multilayer network modules overlap. Non-lazy and hyper-edge similarity walks favour overlap because they lead to longer persistence times among nodes in possibly overlapping groups. 




\bigskip\subsection*{Experiments}.
To illustrate how the network representation affects detected communities in real hypergraphs, we generated a collaboration hypergraph from the 734 references in \emph{Networks beyond pairwise interactions: Structure and dynamics} by F.~Battiston et~al.\cite{battiston2020networks}
We modelled the referenced articles as hyperedges and their authors as nodes. Authors with multiple articles form connections between the hyperedges. We analysed the largest connected component with $|V| = 361$ author nodes in $|E| = 220$ hyperedges.
The median number of authors in a hyperedge is 3, and the authors have contributed to 2.2 articles on average though most have only contributed to one.

We assigned the relative importance of references by their number of citations $c$ in December 2020.
Some references had no citations and some were highly cited.
One such example is \emph{Diffusion of innovations} by Everett M.~Rogers, with more than $120,000$ citations.
To avoid disproportionally large or small hyperedge weights~$\omega(e)$, we weighted the edges by the logarithm of the number of citations and added unit constants to avoid the zero citation problem,
\begin{equation}
\label{eq:citations}
    \omega(e) = \ln\left( c + 1 \right) + 1.
\end{equation}

We modelled the authors' different contributions to articles by assigning higher weights to the first and last author~\cite{chitra2019random}.
We used the edge-dependent node weights
\begin{equation}
\label{eq:contributions}
    \gamma_e(v) = \begin{cases}
        2 \quad \text{if node $v$ is first or last author,} \\
        1 \quad \text{otherwise.}
    \end{cases}
\end{equation}
We assumed equal contribution for alphabetically sorted authors, and assigned all of them weight $\gamma(v) = 1$.
This model ranks a \mbox{co-corresponding} author's contributions lower than those of the corresponding authors.

To study how hypergraph representations and random-walk models affect the community structure, we generated bipartite, unipartite, and multilayer representations for lazy and non-lazy random walks on the collaboration network. We identified nested hierarchical partitions in each network with Infomap, using 100 independent searches for each network. Infomap's running time depends on the number of nodes, links, and solution levels: The bipartite and unipartite representations finished 3--7 times faster than the multilayer representations. The non-lazy bipartite representation with many state nodes ran almost as long.

\begin{table}[tbp]
\caption{Optimised flow-based multilevel communities of the collaboration hypergraph represented with different networks.
The number of nodes includes state nodes for the multilevel representations and the bipartite non-lazy representation.
Shortest codelength of 100 trials with the variance in parenthesis. We measure the overlap as the perplexity of the optimised solutions (see Methods).
\label{table:citations}}
\centering
\setlength{\tabcolsep}{1.8pt}
\begin{small}
\begin{threeparttable}
\begin{tabular}{@{}lrrccccl@{}}
\toprule\noalign{\smallskip}
Representation   & Nodes & Links & \multicolumn{4}{c}{Modules} & \multicolumn{1}{c}{Codelength} \\
{ }              &       &       & Top & Leaf & Levels & Overlap & \multicolumn{1}{c}{(bits)} \\
\noalign{\smallskip}
\midrule\noalign{\smallskip}
\footnotesize{\emph{Lazy}} &&&&& \\
\quad Bipartite                       &   581 &  1,560 & 4 & 23 & 3 & --    & 5.178(1) \\
\quad Unipartite                      &   361 &  2,607 & 9 & 69 & 4 & --    & 3.82557(2) \\
\quad Multilayer                      &   780 & 17,193 & 9 & 76 & 4 & 1.003 & 3.82730(2) \\
\quad Multilayer h-s\tnote{a}         &   780 & 17,193 & 8 & 90 & 4 & 1.127 & 3.54939(3) \\
\noalign{\smallskip} \footnotesize{\emph{Non-lazy}} &&&&&\\ 
\quad Bipartite                       & 1,141 &  3,548 & 5 & 25 & 3 & --    & 5.1733(2) \\
\quad Unipartite                      &   361 &  2,246 & 7 & 49 & 4 & --    & 4.25104(8) \\
\quad Multilayer                      &   780 & 12,843 & 7 & 54 & 4 & 1.098 & 4.16349(8) \\
\quad Multilayer h-s\tnote{a}         &   780 & 12,843 & 9 & 66 & 4 & 1.181 & 3.70432(1) \\
\bottomrule\addlinespace[1ex]
\end{tabular}
\begin{tablenotes}\footnotesize
\item[a] hyperedge-similarity
\end{tablenotes}
\end{threeparttable}
\end{small}
\end{table}

The optimised partitions for the lazy and non-lazy representations behave like the schematic example:
The bipartite representations have the fewest leaf modules and highest codelengths, and the multilayer hyperedge-similarity representations have the most leaf modules and shortest codelengths, with the unipartite and the regular multilayer representations in between (Table~\ref{table:citations}).
Except for the non-lazy bipartite representation with its many state nodes, the lazy representations have more leaf modules and shorter code lengths than their corresponding non-lazy representations because the lazy random walk is more confined than the non-lazy random walk.

With more nodes than in the schematic example, the solutions have more depth.
The bipartite solutions have three, and the unipartite and multilayer solutions have four hierarchical levels. 
The unipartite and multilayer solutions also have more top modules.
With non-lazy dynamics, they split the largest top module, and in the lazy dynamics, they split the two largest top modules. But the second-largest top module reunites in the hyperedge-similarity representation, with stronger connections between similar hyperedges (Fig.~\ref{fig:alluvialCitations} and Fig.~\ref{fig:collaboration-map} in Appendix~\ref{appendix:hypergraph}). The unipartite and multilayer solutions are also most similar at the leaf level (Fig.~\ref{fig:leafmoduleSimilarity} in Appendix~\ref{appendix:similarity}).

\begin{figure}[!tbp]
 \centering
    \includegraphics[width=\columnwidth]{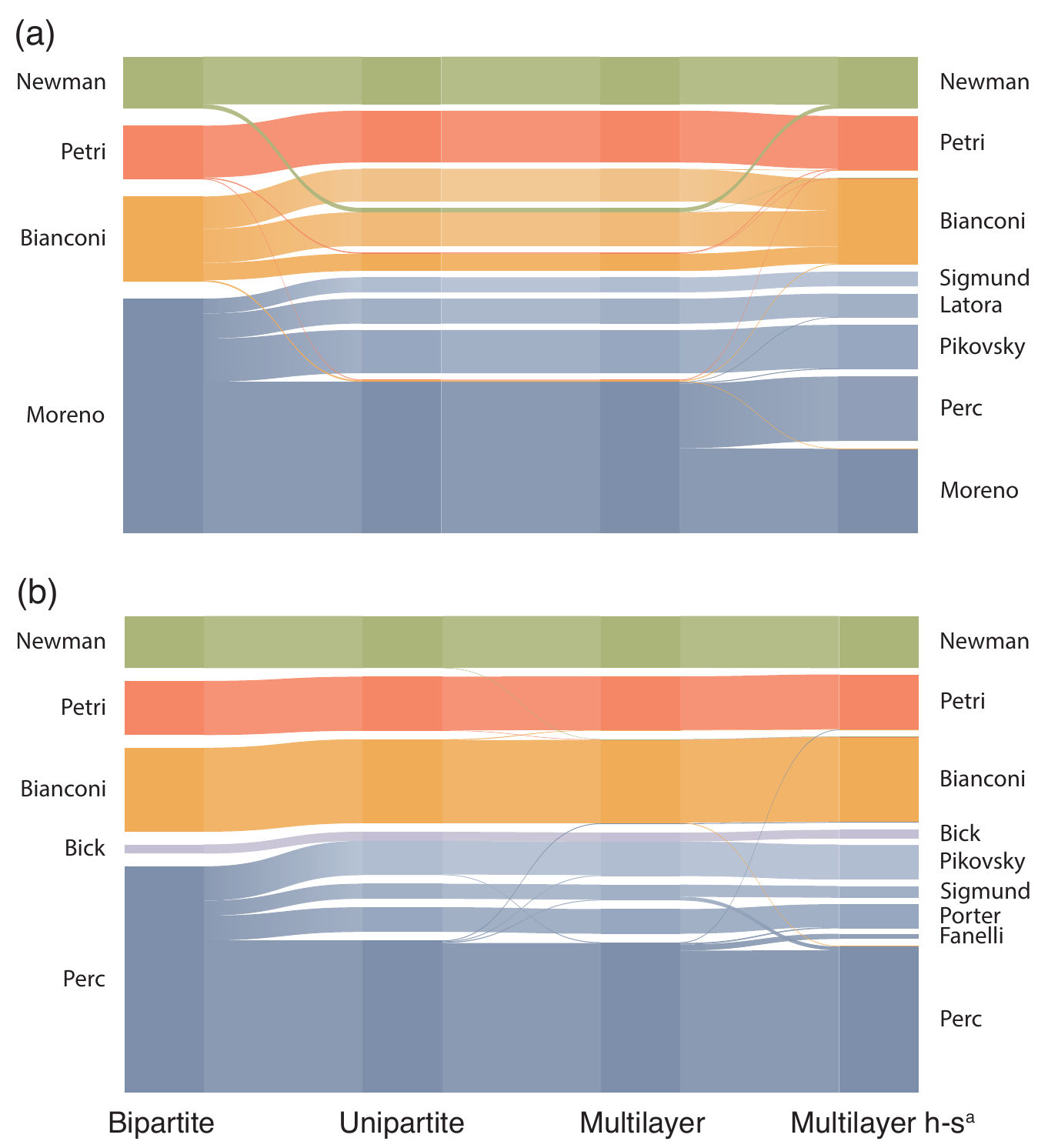}
    \caption{Alluvial diagrams of optimised partitions for different representations of the collaboration hypergraph . Lazy walks in (a) and non-lazy walks in (b). Module names from the top-ranked author within each module.}
    \label{fig:alluvialCitations}
\end{figure}

\begin{figure}[tbhp]
    \centering
    \includegraphics[width=\columnwidth]{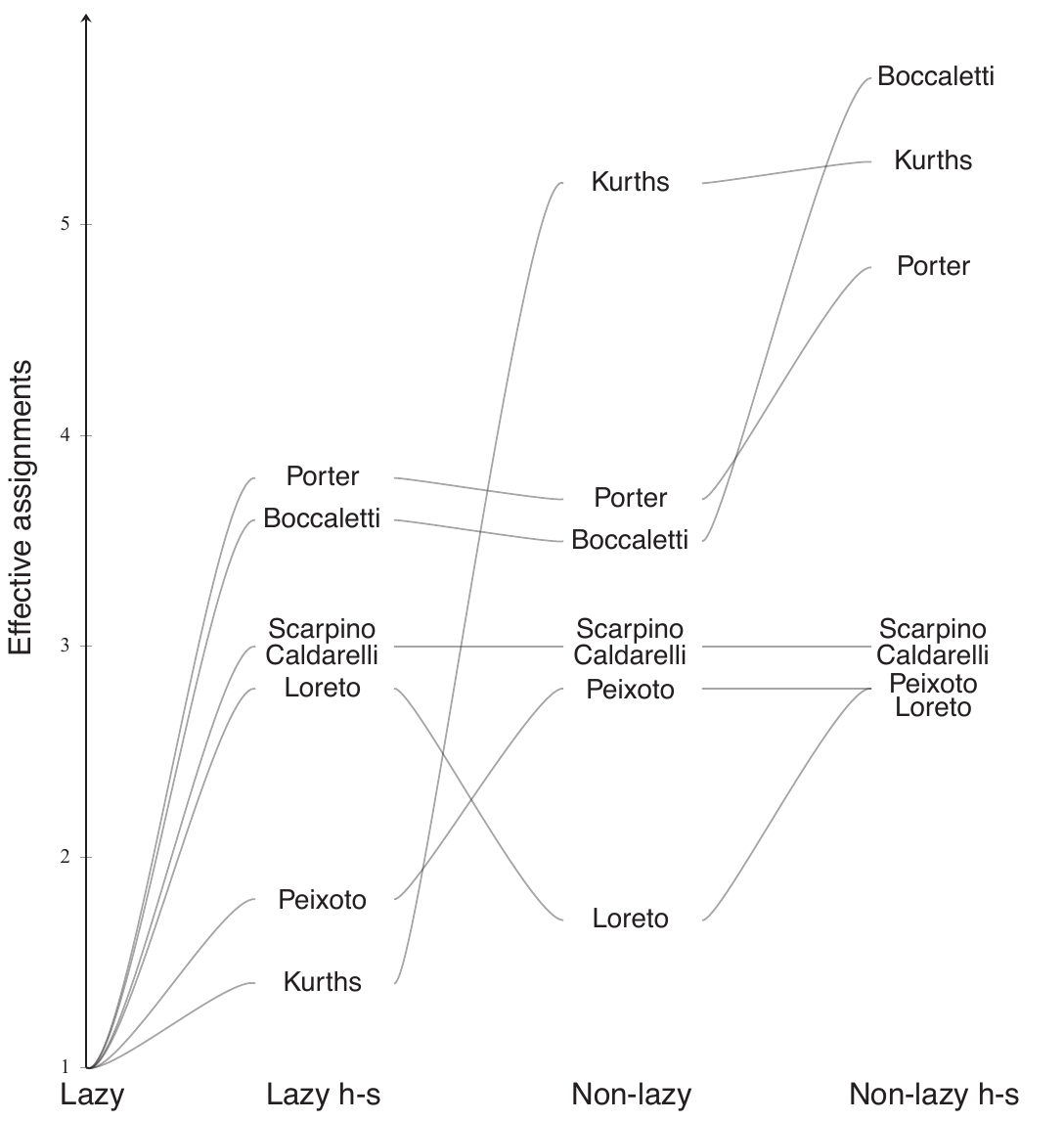}
    \caption{\label{fig:multipleassignedauthors}Authors in the collaboration hypergraph with the highest average effective number of assignments in the lazy and non-lazy multilayer representations (see Methods).}
\end{figure}

In this larger example, the multilayer hyperedge-similarity representations give more overlap.
The non-lazy representations result in higher average overlap because random walkers visiting a node must continue to other nodes, often in the same or a similar hyperedge layer.
When random walkers from dissimilar hyperedges come together at a node, they tend to return to where they came from and favour overlapping modules.
The non-lazy representations also result in higher max overlap with the same authors topping all representations (Fig.~\ref{fig:multipleassignedauthors}).

In line with the information-theoretic duality between finding regularities in data and compressing those data, representations that enable deeper solutions with more modules have shorter codelengths (Table \ref{table:citations}).
The lazy multilayer representation is an exception.
Its optimised codelength is bound above by the lazy unipartite representation's codelength -- they have the same codelength for the same hard partition -- and overlapping modules can potentially reduce the codelength. Infomap's best codelength was instead 0.05 percent longer than for the lazy unipartite representation. Multilayer representations with their many state nodes and links aggravate the search problem, and Infomap could not find a better solution in 100 attempts. But the gain from overlapping modules is higher for the non-lazy multilayer representation and Infomap finds a solution with a significantly shorter codelength.

\bigskip\subsection*{A case study on fossil data}.
Palaeontologists classify major groups of marine animals archived in the fossil record into global-scale faunas that change over time\cite{sepkoski_factor_1981}. They have used different network representations to understand the macroevolutionary pattern of marine biodiversity\cite{rojas_multiscale_2019, muscente_quantifying_2018}. However, it is still unclear how such an organisation of marine animals into modules representing global faunas changes with random-walk model and network representation. 
To illustrate how the network representation of the underlying paleontological data affects empirical estimates of this macroevolutionary pattern, we generated a hypergraph from genus-level fossil occurrences presented in ref.~\citen{rojas_multiscale_2019} and retrieved from the PaleoDB\cite{peters_paleobiology_2016}. We restricted our analysis to fossil occurrences from the Cambrian (541 MY) to the Cretaceous period (66 MY) and modelled 77 geological stages as hyperedges and 13,276 genera as nodes. Genera occurring in multiple geological stages form connections between hyperedges. We weighted the hyperedges by dividing the number of samples where a genus occurs in a given geological stage by the total number of samples recorded at the stage, a procedure modified from ref.~\citen{rojas_global_2017}. We generated bipartite, unipartite, and multilayer network representations for lazy and non-lazy random walks from the underlying palaeontology data and identified optimised partitions in the assembled networks using Infomap. 

\begin{figure}[htb]
 \centering
    \includegraphics[width=\columnwidth]{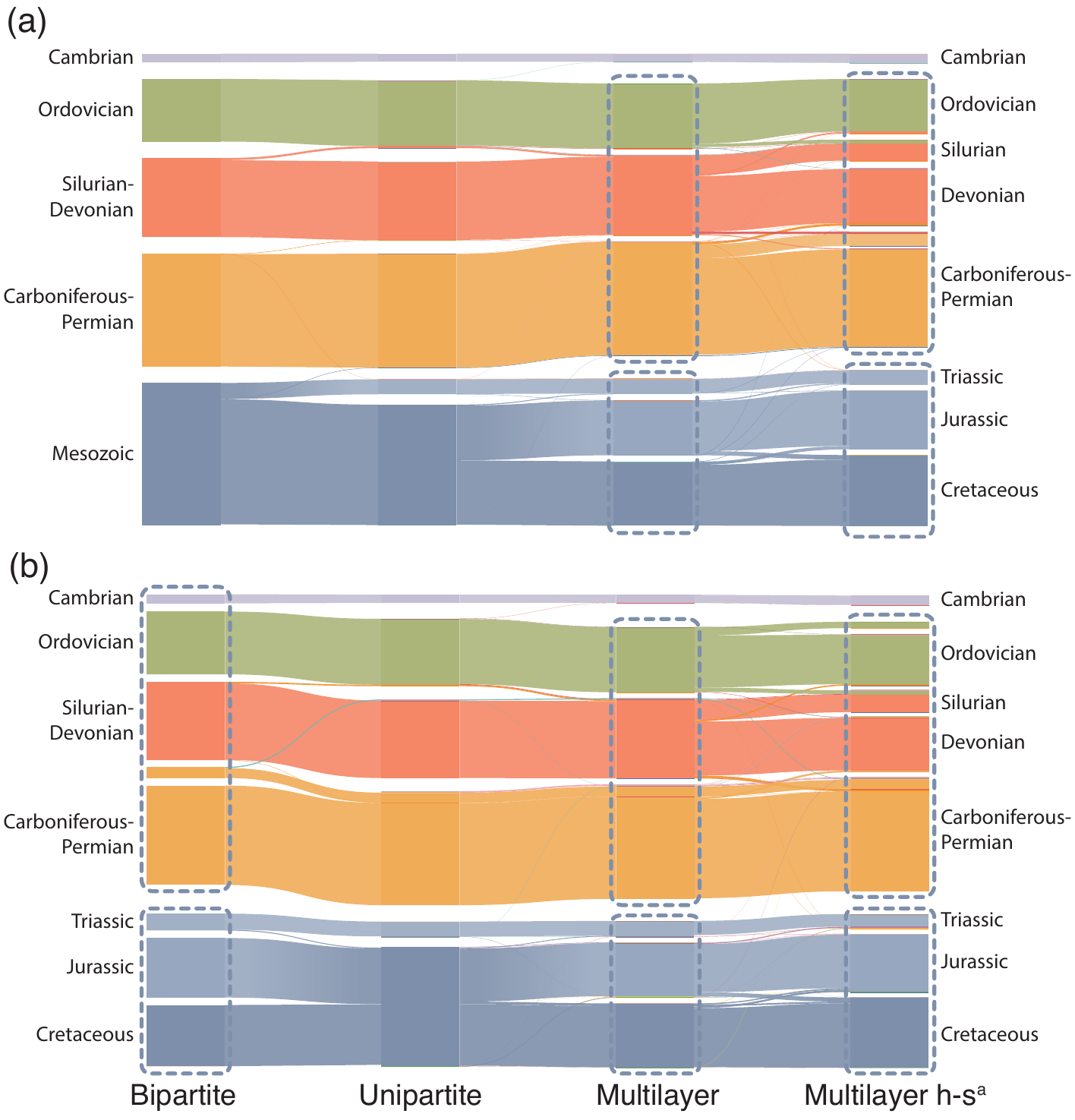}
    \caption{Alluvial diagrams of optimised partitions for the fossil hypergraph represented with different networks. Lazy walks in (a) and non-lazy walks in (b). We show top modules when a partition lacks deeper levels and leaf modules marked with dashed lines when they exist. Module names from the geological period or era represented by the fauna assemblage.
    \label{fig:fossilalluvial}}
\end{figure}

For lazy random walks, Infomap partitioned only the multilayer representations into multilevel communities: three modules at the first hierarchical level [Fig. \ref{fig:fossilalluvial}(a)]. Similar to the schematic example and the collaboration hypergraph, the bipartite representation for the lazy random walks has the fewest leaf modules and the highest codelength. The multilayer hyperedge-similarity representation has the most leaf modules and the shortest codelength (Table \ref{table:fossil}).

\begin{table*}[tbp]
\caption{Optimised flow-based multilevel communities of the fossil hypergraph represented with different networks.
The number of nodes includes state nodes for the multilevel representations and the bipartite non-lazy representation.
The number of non-trivial top and leaf modules. Average number of levels weighted by the flow volume. We measure the overlap as the perplexity of the optimised solutions (see Methods).
Shortest codelength of 20 trials with the variance in parenthesis.
\label{table:fossil}}
\centering
\setlength{\tabcolsep}{6pt}
\begin{small}
\begin{threeparttable}
\begin{tabular}{@{}lcrcccclc@{}}
\toprule\noalign{\smallskip}
Representation   & Nodes & \multicolumn{1}{c}{Links} & \multicolumn{4}{c}{Modules} & \multicolumn{1}{c}{Codelength} & Time      \\
{ }              & $(\times 10^3)$ & $(\times 10^3)$ & Top  & Leaf  & Levels  & Overlap & \multicolumn{1}{c}{(bits)} & (hh:mm:ss) \\
\noalign{\smallskip}
\midrule\noalign{\smallskip}
\footnotesize{\emph{Lazy}} &&&&& \\

\quad Bipartite                       & 13 &      79 & 5 &  8 & 2.02 & --    & 10.50927(5) & 00:00:06 \\
\quad Unipartite                      & 13 &  16,155 & 6 & 13 & 2.02 & --    & 10.3953503(1) & 00:13:24 \\
\quad Multilayer                      & 40 & 174,490 & 3 & 17 & 3.00 & 1.011 & 10.39819(1) & 09:08:43 \\
\quad Multilayer h-s\tnote{a}         & 40 & 174,490 & 3 & 19 & 3.28 & 1.135 & \,\,\,9.84170(1) & 14:19:39 \\

\noalign{\smallskip} \footnotesize{\emph{Non-lazy}} &&&&&\\ 
\quad Bipartite                       & 53 &  25,937 & 2 & 15 & 3.02 & --    & 10.34889(3) & 01:14:25  \\
\quad Unipartite                      & 13 &  16,141 & 6 & 12 & 2.02 & --    & 10.4031798(6) & 00:13:04 \\
\quad Multilayer                      & 40 & 174,209 & 3 & 15 & 3.00 & 1.010 & 10.406141(9) & 08:55:03 \\
\quad Multilayer h-s\tnote{a}         & 40 & 174,209 & 3 & 16 & 3.00 & 1.135 & \,\,\,9.84912(1) & 13:23:13 \\

\bottomrule\addlinespace[1ex]
\end{tabular}
\begin{tablenotes}\footnotesize
\item[a] hyperedge-similarity
\end{tablenotes}
\end{threeparttable}
\end{small}
\end{table*}

For non-lazy random walks, Infomap partitioned the bipartite representation into a multilevel solution with shorter codelength than the unipartite representation and the standard multilevel representation [Fig.~\ref{fig:fossilalluvial}(b)]. The multilayer hyperedge-similarity representation once more provides the most leaf modules and the highest overlap. 

The multilayer network representations, including lazy and non-lazy random walks, reproduce modules reminiscent of the Cambrian, Paleozoic, and modern evolutionary faunas widely used in macroevolutionary research\cite{sepkoski_factor_1981}. Also, leaf modules in the multilayer representations capture subfaunas from specific geological periods as nested modules such as Silurian, Triassic, Jurassic, and Cretaceous. Infomap applied to the bipartite representation of the non-lazy random walks identified similar subfaunas but combined Cambrian and Paleozoic faunas into a single top module, obscuring the large-scale pattern. Overall, our results indicate some advantages of using multilayer over bipartite and unipartite representations of fossil occurrence data to quantify the marine biodiversity's macroevolutionary patterns, with lazy and non-lazy random walks providing similar solutions.

\section*{Conclusions}
We have derived unipartite, bipartite, and multilayer network representations of hypergraph flows with different advantages.
We used the information-theoretic and flow-based community detection method Infomap to explore how different hypergraph random-walk models and network representation change the number, size, depth, and overlap of identified multilevel communities.
By identifying flow-based communities both in a schematic and real hypergraphs -- a small collaboration hypergraph of researchers working on networks beyond pairwise interactions and a large faunal hypergraph of sampled species across geological stages -- we found that the bipartite network representation is the most compact and enables the fastest community detection.
A multilayer network representation that reinforces flows within similar layers -- one for each hyperedge -- gave the deepest modular structures with the most module overlap.
But the modular detection gain comes at a high computational cost: Combining fully connected layers with other layers requires many more nodes and links than in the bipartite network representation.
If the research question does not require hyperedge assignments or overlapping modules, the unipartite network representation provides a trade-off with intermediate compactness, speed, and the ability to reveal modular regularities.
Among the random-walk models, lazy walks typically give more modules in deeper nested structures, and non-lazy walks provide higher modular overlap.
Our methods and results help researchers model and map flows on hypergraphs to study the effects of multibody interactions in complex systems.




\section*{Methods}

\subsection*{Unrecorded teleportation}. With hyperedge-independent node weights where $\gamma_e(u) = \gamma(u)$ for all hyperedges $e \in E(u)$, undirected weighted networks can represent the dynamics, and the stationary distribution of the random walk $\pi_u$ is proportional to the product of node $u$'s total incident hyperedge weight $d(u)$ and weight $\gamma(u)$.
With normalised node-visit rates\cite{chitra2019random},
\begin{align}\label{eq:stat_dist_node}
    \pi_u = \frac{d(u)\gamma(u)}{\sum_{v \in V}d(v)\gamma(v)}. 
\end{align}
For the multilayer network representation, the node-visit rates split between layers based on the node $u$'s incident hyperedge weight per layer state node
\begin{align}\label{eq:stat_dist_statenode}
    \pi_{u}^{\layerE} = \frac{\omega(\layerE)\gamma(u)}{\sum_{v \in V}d(v)\gamma(v)}. 
\end{align}

With hyperedge-dependent node weights $\gamma_e(u)$, only directed weighted networks can represent the dynamics.
We use random teleportation to ensure ergodic walks when deriving the node-visit rates with the power-iteration method.
Unrecorded teleportation to links minimises the distortion\cite{lambiotte2012ranking}:
In each iteration of the power-iteration method, we distribute a fraction $\tau=0.15$ of each node's flow volume among all nodes proportional to their out-link weights.
The remaining flow volume moves on the links proportional to their weights.
In the last iteration, we move all flows on the links proportional to their weights and record all flows on links and nodes to obtain the ergodic node- and link-visit rates with unrecorded teleportation.
This procedure gives equivalent visit rates as simulating a random walker that only records moves on links:
With probability $1-\tau$, the random walker moves to a node by following the links proportional to their weights and records the link and the target node.
With probability $\tau$, the random walker teleports without recording to the link's source node proportional to the link weight.
The normalised number of recordings of each node and link gives the visit rates.

We want teleportation applied to undirected networks -- where it is unnecessary -- to leave the node- and link-visit rates unchanged.
We achieve this smooth teleportation by scaling the transition rates from nodes by the node-visit rates:
Then unrecorded teleportation proportional to the nodes' total out-link weights followed by recorded moves on the links proportional to their weights distributes on the nodes according to the ergodic visit rates on undirected networks\cite{lambiotte2012ranking}.
For the general case when the node weights can depend on the hyperedge, and the network may be directed, we use Eq.~\ref{eq:stat_dist_node} without assuming $\gamma_e(u) = \gamma(u)$ as an approximation of the node-visit rates:
\begin{align}
    \tilde{\pi}_u = \frac{\sum_{e \in E(u)}\omega(e)\gamma_e(u)}{\sum_{v \in V\!\!, e \in E(v)}\omega(e)\gamma_e(v)} 
\end{align}
for nodes and
\begin{align}
    \tilde{\pi}_{u}^{\alpha} = \frac{\omega(\layerE)\gamma_\layerE(u)}{\sum_{v \in V\!\!, e \in E(v)}\omega(e)\gamma_e(v)}\text{ for } \layerE \in E(u)
\end{align}
for state nodes.
With exact node-visit rates, we would obtain the stationary flow volumes on links by multiplying the transition rates by the source nodes' visit rates. With approximate node-visit rates, instead, we obtain the link weights 
\begin{align}
    w_{ue} = \tilde{\pi}_u P_{ue}
\end{align}
for bipartite networks,
\begin{align}
    w_{uv} = \tilde{\pi}_u P_{uv}
\end{align}
for unipartite networks, and 
\begin{align}
    w_{uv}^{\layerE\layerF} = \tilde{\pi}_{u}^{\alpha} P_{uv}^{\layerE\layerF}\text{ for } \layerF \in E(u,v)
\end{align}
for multilayer networks.
With unrecorded teleportation proportional to these link weights, modelling flows on hypergraphs give node-visit rates robust to changes in the teleportation rate and independent of the representation.

\bigskip\subsection*{Overlap metric}. Modules overlap when Infomap assigns a node's state nodes in the multilayer network representations to different modules. Measuring the overlap through the absolute number of assignments is misleading because the overlap is 2 regardless of the number of state nodes assigned to a different module than the rest.
Instead, we used the effective number of assignments.
If a fraction~$f$ of node~$u$'s state nodes is assigned to the $m$th module in $u$'s module assignment set, the $m$th element of $u$'s assignment vector is $a^u_m = f$ and the effective number of assignments measured by the perplexity of $u$'s module assignments is
\begin{equation}
    o_u = 2^{H(\mathbf{a}^u)}.
\end{equation}
The effective number of assignments is one if all $u$'s state nodes are in one module, and it is equal to the number of assignments when the state nodes are divided evenly among $u$'s module assignments.
We averaged over all nodes for the partition overlap.


\section*{Data and code availability}

All data and source code are available on GitHub: \small{\url{http://github.com/mapequation/mapping-hypergraphs}}.

\bibliographystyle{naturemag}
\bibliography{references.bib}

\begin{acknowledgments}

We thank Christopher Bl\"ocker, Manlio De Domenico, Michael Schaub, and Jelena Smiljani\'c for valuable comments that helped us improve the manuscript. A.E was supported by the Swedish Foundation for Strategic Research, Grant No.\ SB16-0089. A.R., D.E.\ and M.R.\ were supported by the Swedish Research Council, Grant No.\ 2016-00796. 

The computations was enabled by resources provided by the Swedish National Infrastructure for Computing (SNIC) at High Performance Computing Center North (HPC2N), partially funded by the Swedish Research Council through grant agreement no.\ 2018-05973.

\end{acknowledgments}

\section*{Author contributions}
A.E.\ and M.R.\ conceived the study. A.E., A.R.\ and D.E.\ performed the numerical experiments and analysed the results. A.E.\ and M.R.\ wrote the manuscript.

\section*{Competing interests}

The authors declare no competing interests.


\cleardoublepage
\onecolumngrid
\appendix

\section{Appendix}
\label{appendix:hypergraph}
\label{appendix:similarity}

\begin{figure*}[htb]
    \centering
    \includegraphics[width=0.80\textwidth]{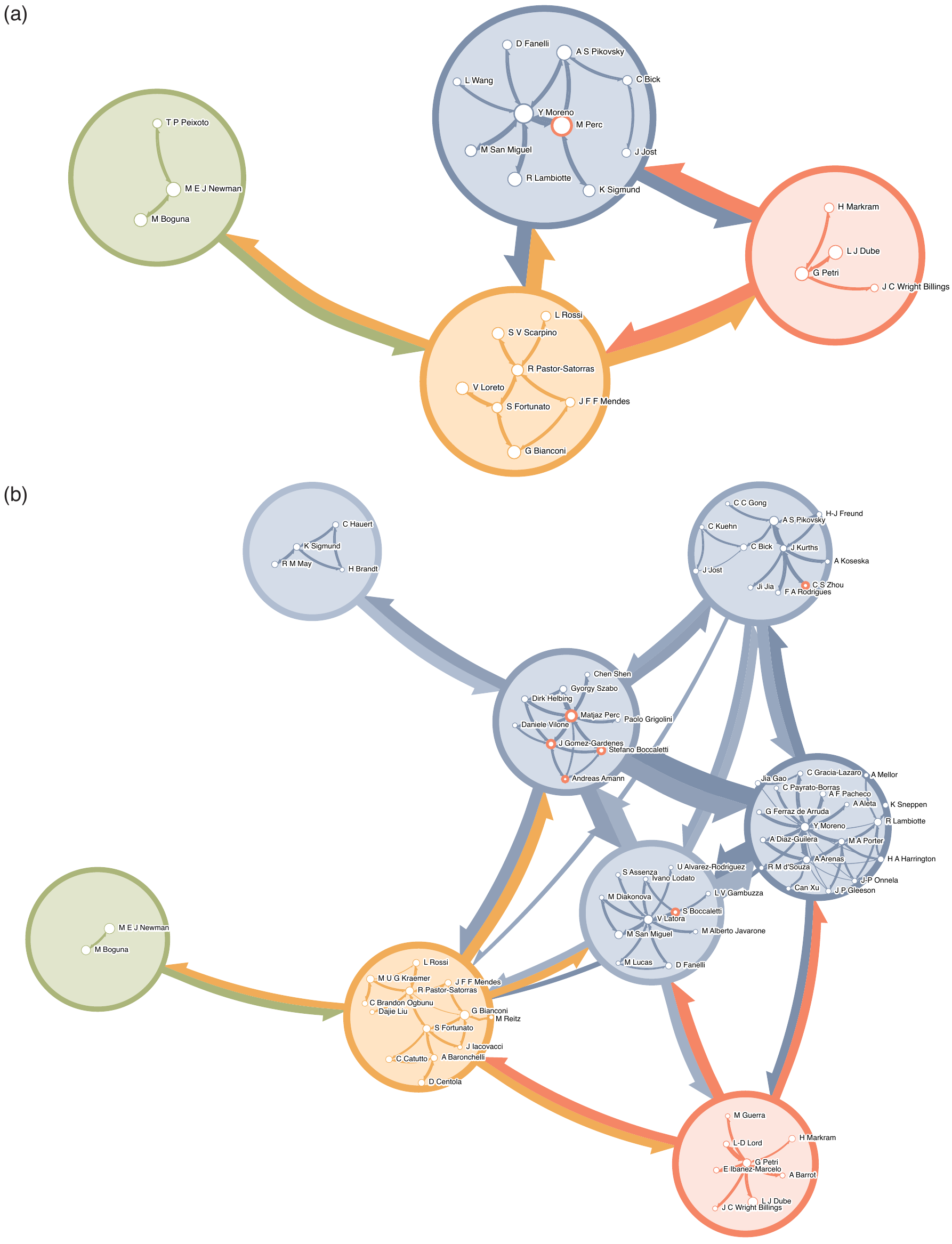}
    \caption{Hierarchical maps of the collaboration hypergraph using
    (a) the bipartite representation and
    (b) the multilayer hyperedge-similarity representation.
    Module colours are the same as in Fig.~\ref{fig:alluvialCitations}(a).
    Aggregated inter-module links with sizes proportional to the exiting flow volume and length inversely proportional to the flow volume.
    White sub-modules are labelled with the top-ranked author.
    The largest blue top module in (a) contains ten sub-modules. In (b), the partition assigns those nodes to five top modules containing more sub-modules.
    S.~Boccaletti, one of the most overlapping authors and highlighted in red, is assigned to one module in (a) and three top modules and six sub-modules in (b).}
    \label{fig:collaboration-map}
\end{figure*}

\clearpage

\begin{figure}[htpb]
    \centering
    \includegraphics[width=0.6\textwidth]{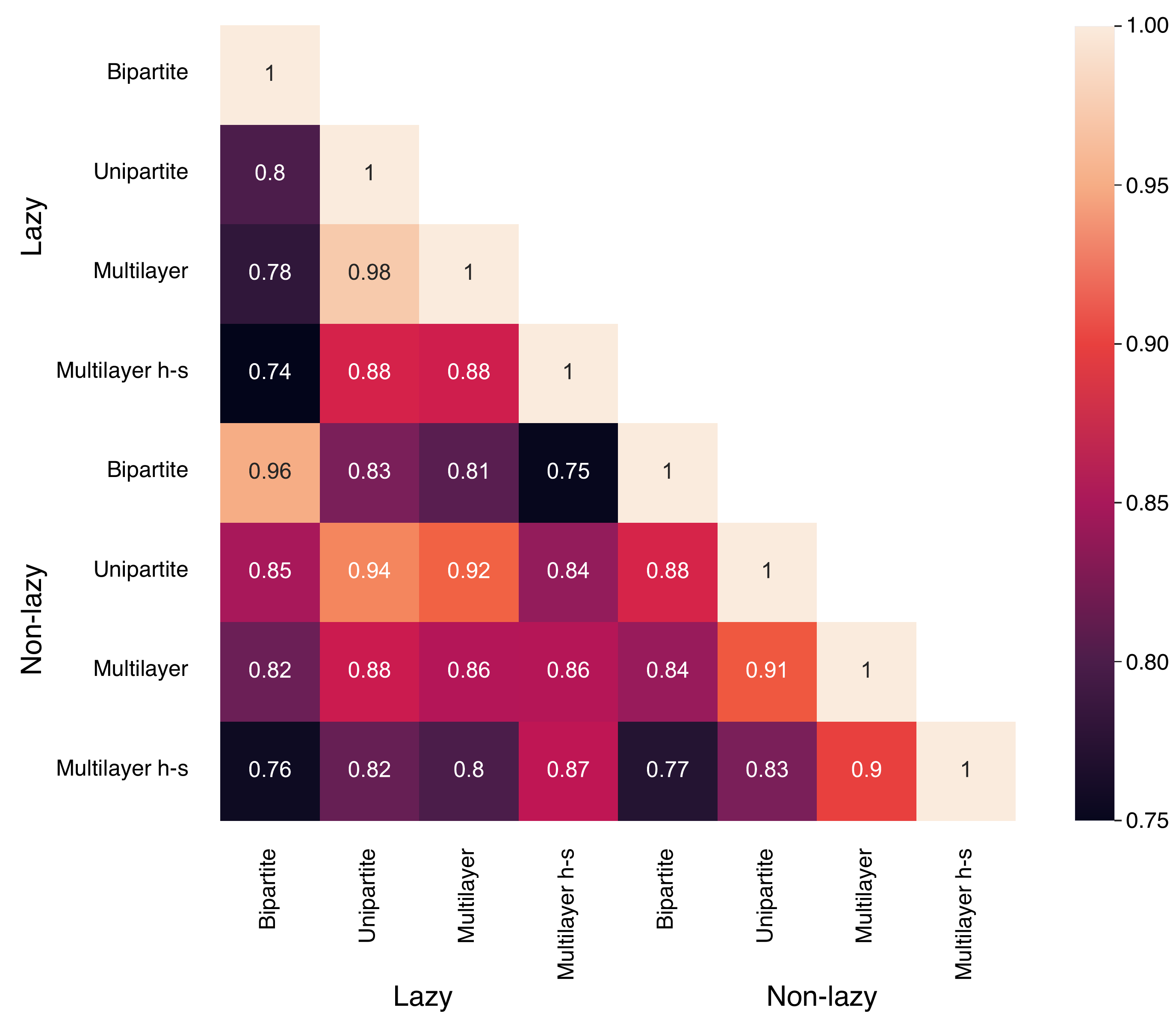}
    \caption{Leaf module assignments' adjusted mutual information for different random walk dynamics and hypergraph representations. The bipartite representations differ the most from the other representations, and the unipartite and multilayer representations are most similar at the leaf level.}
    \label{fig:leafmoduleSimilarity}
\end{figure}

\end{document}